\documentstyle[preprint,aps]{revtex}

\begin{document}
\title{Molecular vibration in cold collision theory}
\author{Alessandro Volpi and John L. Bohn \cite{byline}}
\address{JILA and Department of Physics, University of Colorado, Boulder, CO}
\date{\today}
\maketitle

\begin{abstract}

Cold collisions of ground state oxygen molecules with
Helium have been investigated in a wide range of cold
collision energies (from 1 $\mu$K up to 10 K) treating
the oxygen molecule first as a rigid rotor and then 
introducing the vibrational degree of freedom.
The comparison between the two models shows that
at low energies the rigid rotor approximation
is very accurate and able to describe all the dynamical
features of the system.
The comparison
between the two models has also been extended to cases
where the interaction potential He - O$_2$ is made
artificially stronger. 
In this case vibration can perturb rate constants,
but fine-tuning the rigid rotor potential can
alleviate the discrepancies between the two models.

\end{abstract}


\narrowtext

A wide variety of techniques is now available 
for the production of ultracold molecular gases, and 
collisions often play a crucial role in their success. 
The challenge for theoreticians is then a detailed understanding
of the scattering properties at low temperatures, in order to
assess the feasibility of the experiments.
In this paper, our aim is to investigate the role of molecular
vibration in cold collisions, a topic that has been already 
object of interest for instance in the case of
molecules produced by photoassociation (PA)
\cite{Fioretti,Knize,Stwalley,Heinzen}.
This technique exploits one or more PA lasers
to catalyze alkali
atom pairs into bound molecular states \cite{Paul},
resulting in molecules with nearly the same
translational temperature as the original atomic
sample ($\sim$ 100 nK
in the case of extraction from a Bose-Einstein 
condensate \cite{Heinzen}).
Although rotationally cold, these molecules are typically formed in high
vibrational states, from which they can relax by collisions with
another free atom. The understanding and characterization
of this relaxation process
is crucial, because a large amount of energy is released in the
process, and this quenching can dramatically 
affect the efficiency of the cooling \cite{Forrey}.
A study in this sense has been conducted for the
He - O$_2$ system \cite{baldalJPCA01},
pointing out 
the importance of understanding the role
of vibration in ultracold molecular collision.

The buffer-gas cooling (BGC) technique \cite{Doylemag,DoyleVO}, 
in which molecules are cooled by collisions with a cold buffer
gas, also demands a detailed picture of the collisions.
The lowest temperature reached so far is 0.3 K in a gas
of CaH molecules, a result achieved by Doyle's group \cite{DoyleCaH}
 who used the
same technique recently to cool PbO molecules \cite{egoweipatfridoyPRA01}. 
Appealing candidates for BGC would be O$_2$ molecules, already investigated
from the theoretical point of 
view \cite{friJCSFT98,bohPRA00,avdbohPRA01},
and NH, for which in addition magnetic trapping
has been proposed recently by the alternative
Stark slowing technique \cite{vanjonbetmeiPRA01}.

Lower temperatures can be achieved via 
evaporative cooling, in which high-energy molecules
are selectively removed from the trap.
The success of
the cooling requires that elastic 
rethermalizing collisions
be much more efficient than the spin-changing process
that produces untrapped molecules.
Since several of our previous papers have demonstrated that
spin orientations can be easily 
changed in cold molecular collisions,
an important question raised in this work
is whether the vibrational motion
can further encourage spins to flip, 
even in the ground vibrational state
of the colliding molecule.

Following Refs. \cite{bohPRA00,avdbohPRA01,volbohPRA02},
we evaluate here the state-to-state rate constants for 
collisions of molecular oxygen
($^{17}$O$_2$ isotopomers) with $^{3}$He atoms, 
using a recently published potential
energy surface (PES) \cite{grostrJCP00}.
We first treat O$_2$ as a rigid rotor, then
we introduce the vibrational degree of freedom
in order to assess the influence
of the vibrational motion.
Results can also be compared with a previous calculation\cite{bohPRA00}
on the same system performed by one of the authors
on a different PES \cite{Cyb} using a rigid rotor model.
We show how
the rigid rotor approximation holds
quite well for this system, as expected.
This is because the 
atom-molecule interaction is very weak compared to the
O - O interaction, and the O$_2$ molecule
is strongly bound and physically
very similar to a rigid rotor. 

Nevertheless,
for stronger three-body interactions or for
shallower diatomic potentials, the rigid rotor
model could be unrealistic in describing the
dynamical features of the system. What might happen
in such a case is as well a subject of this paper.
To this end, we consider a set of
artificial models, by increasing the atom -
diatom interaction strength while preserving 
the natural O $-$ O interaction. 
Thus at some point the He atom in the 
artificial model will gain sufficient
kinetic energy during the collision
to excite vibrations of the O$_2$ molecule.
The effect of vibrational degrees of freedom
is primarily to provide additional phase shifts,
which may be "mocked up" by suitable
adjustment to the short-range potentials.
In addition, vibrational resonances are
introduced, but these should be small in number
as compared to rotational resonances.

Throughout this paper
we report energies in units of Kelvin by dividing by the Boltzmann
constant $k_B$.  These units are related to the more familiar
wave numbers via 1 K = 0.695 cm$^{-1}$. Lengths are expressed
in units of the bohr radius $a_0$.

We will consider molecules consisting of two
$^{17}$O atoms, whose nuclear spin $i$ is equal to $5/2$. We
assume that total spin
{\bf I} = {\bf i}$_1$ + {\bf i}$_2$ is conserved in the collision,
implying that the even molecular
rotational states $N$ are separated from the odd ones
\cite{Mizushima}. Following 
Refs. \cite{bohPRA00,avdbohPRA01,volbohPRA02},
we limit our discussions to the ``even-$N$'' manifold
of molecular states.
However, we have performed some sample 
calculation for the ``odd-$N$'' manifold, finding results
consistent with the conclusions
for the case of even $N$.

Including the $S$ = 1 electronic spin of
O$_2$, at low energies the molecules have
total spin {\bf J} = {\bf N} + {\bf S} = 1. 
The Hamiltonian operator ${\bf \hat{H}}_{{\rm O}_2}$ 
for molecular oxygen is given by
\begin{equation}
\label{O2Ham}
{\bf \hat{H}}_{{\rm O}_2} = -{\hbar^2 \over 2 \mu_{{\rm O}_2}} 
\left[ {d^2 \over dr^2}
-{ N(N+1) \over r^2} \right] + V(r) + {\bf \hat{H}}_{fs}
\end{equation}
where $\mu_{{\rm O}_2}$ is the O - O reduced mass
and $V$ is the atom-atom potential
depending on the stretching coordinate $r$. We have taken
the fine-structure Hamiltonian ${\bf \hat{H}}_{fs}$ from 
Ref. \cite{fremildeslurJCP70} disregarding
the molecular hyperfine interaction and using
the fine structure parameters determined in \cite{Cazzoli}
by microwave spectroscopy. We assume these parameters
to be the same for the ground and excited
vibrational states.

The Hamiltonian in (\ref{O2Ham}) refers to the vibrating
diatom model, and obvious simplifications lead it to the appropriate
expression for the rigid rotor.
We note here that the molecular rotational quantum number $N$
is no longer strictly a good quantum number for the molecular states,
because different values of $N$ are 
coupled together by the fine-structure
Hamiltonian ${\bf \hat{H}}_{fs}$ (see eq. (A5) in Ref.
\cite{fremildeslurJCP70}). However, since
the fine-structure coupling is small compared to the
rotational separation, different $N$'s are mixed
only weakly, so we will continue to 
use $N$ to label the true channels in what follows.

As in Ref. \cite{bohPRA00} we will focus our attention 
on the weak-field-seeking state 
$|N\;J\;M_J \rangle$ = $|0\;1\;1 \rangle$ of the molecule.
We are concerned both with the elastic collisions that 
preserve this state, and with "loss" collisions that 
yield the untrapped states
$|0\;1\;0 \rangle$ and $|0\;1\;-1 \rangle$. 
The relevance of
these collisions for ultracold
molecular studies has already been discussed
in \cite{bohPRA00,avdbohPRA01,volbohPRA02}.

Ref. \cite{bohPRA00} provided the theoretical
framework for atom - diatom scattering,
along the lines of the model originally due to Arthurs
and Dalgarno \cite{Arthurs,Child}, and properly modified
to incorporate the electronic spin of the oxygen molecule.
Here we briefly recall some features needed to understand
the present calculation, pointing out the differences
between the rigid rotor and the vibrating diatom approaches.
We compute rate constants in zero magnetic field, as we are interested
primarily in the comparison of the rate constants
between the two models.
The study of molecular collisions in a magnetic
field is however a central topic for trapping purposes,
which we have discussed elsewhere \cite{volbohPRA02}.

After multiplying the wave function by $R$ in order 
to remove first derivatives,
the full Hamiltonian operator describing the He - O$_2$ collision is 
given by
\begin{equation}
\label{totalH}
{\bf \hat{H}} = -{\hbar^2 \over 2 \mu} \left[ {d^2 \over dR^2}
-{ {\bf \hat{L}}^2 \over R^2} \right] + {\bf \hat{H}}_{{\rm O}_2} + V'(R,r,\theta)
\end{equation}
where $\mu$ is the reduced mass for the He-O$_2$ system,
$R$ is the length of the Jacobi vector joining the atom
to the molecule center-of-mass, ${\bf \hat{L}}^2$ is the
centrifugal angular momentum operator and ${\bf \hat{H}}_{{\rm O}_2}$
is the molecular oxygen Hamiltonian defined in (\ref{O2Ham}).
The potential term $V'$, depending in general on $R$, $r$, and
the bending angle $\theta$ that the molecule's axis makes
with respect to ${\bf R}$, accounts for the
He - O$_2$ interaction \cite{grostrJCP00}.
The O - O interatomic contribution is
already included in the molecular Hamiltonian ${\bf \hat{H}}_{{\rm O}_2}$.
We note here that the PES considered in this work differs
from the one by Cybulski et al. \cite{Cyb} used in a previous calculation 
by one of the authors \cite{bohPRA00}.
In particular, the potential well for the three-body interaction
is found to be about 30\% deeper for the new PES with respect to
the preceding one (however, the PES \cite{Cyb} was
deepened by 20\% in the calculations presented in \cite{bohPRA00}).
The two potential surfaces provide consistent 
results: the $s$-wave scattering 
lengths are $\sim$ $- 2.9$ a.u. and $\sim$ $- 1.5$ a.u.
using the PES \cite{grostrJCP00} and \cite{Cyb}, respectively.

If the rigid rotor model is adopted, the internuclear
distance $r$ is frozen to its equilibrium value $r_0 =$
2.282 $a_0$ and any dependence of the Hamiltonian on this coordinate
is neglected. In this case, the theoretical formulation
of the problem reduces to the one in \cite{bohPRA00},
to which we refer for details. 
For the vibrating diatom, the $r$-dependence of the Hamiltonian
is taken into account.

The full multichannel calculation requires casting $V'(R,r,\theta)$
in an appropriate angular momentum basis. 
Namely, we express the
Hamiltonian in a basis of total angular momentum 
${\cal J} = {\bf N} + {\bf S} + {\bf L}$, 
in terms of the molecule's mechanical rotation (${\bf N}$),
its electronic spin (${\bf S}$), and the partial wave
representing the rotation of the molecule and the He atom
about their center of mass (${\bf L}$). 
Our basis for close-coupling calculations is then
\begin{equation}
\label{basis}
|{\rm O}_2(^3\Sigma_g^-) \rangle |{\rm He}(^1S) \rangle
| v N [ J L ] {\cal JM} \rangle
\end{equation}
where the electronic spin quantum number $S$ 
is not explicitly indicated being always
equal to 1 in the calculation presented here.
Vibrational wave functions are computed
for a particular value of $N$ then transformed to
the $J$ basis.
Evaluation of $V'$ in the basis (\ref{basis}) 
has been discussed in Ref. \cite{bohPRA00}. As for
the integration of the matrix elements
over the vibrational coordinate $r$, we have
performed numerical Gaussian quadratures.

Once the Hamiltonian is in place the coupled-channel
equations are solved subject to scattering boundary
conditions to yield scattering matrices.  Since we assume
zero magnetic field the total
angular momentum ${\cal J}$ is a good quantum number, and
moreover the results are independent of the laboratory
projection ${\cal M}$ of total angular momentum. In
the context of magnetic trapping, the 
resulting total-${\cal J}$ scattering matrices can be 
conveniently transformed to a basis labeling the magnetic
quantum numbers, namely 
$\langle v N JM_J LM_L |S| v^{\prime} N^{\prime}
J^{\prime}M_J^{\prime} L^{\prime}M_L^{\prime} \rangle$.
Note that in general all the quantum numbers $v$, $N$, $J$, $M_J$,
$L$, and $M_L$ are subject to change in a collision.  
However, at the energies we consider, only changes in $M_J$
are energetically allowed.
Cross sections and state-to-state rate coefficients 
are then obtained as in \cite{bohPRA00}.

For $^3$He - $^{17}$O$_2$ collisions,
rate constants relative to the entrance
channel $| N\;J\;M_J \rangle$ = $| 0\;1\;1 \rangle$
for the elastic transition and the inelastic ones
(to the states 
$| 0\;1\;-1 \rangle$ and $| 0\;1\;0 \rangle$)
have been calculated in a wide range of collision
energy, from 1 $\mu$K up to 10 K, using
both the rigid rotor and the vibrating diatom model.
We have found
that rotational states up to $N$ = 8 and partial
waves up to $L$ = 8 must be retained in the calculations.
Scattering calculations are performed
using a log-derivative
propagator method \cite{johJCP73}
starting from $R$ = 4.1 bohr. We separate the
propagation into two parts, from $R$ = 4.1 to 24.0 bohr,
with a step size of 0.01 bohr, and then from
$R$ = 24.0 until the asymptotic limit of $R_{max}$ = 450 bohr
adopting a larger step size of 0.1 bohr.
These parameters assure rate constants convergent
within less than 1\%.

We recall the rovibrational structure of the 
oxygen molecule. The zero-point energy is $\sim$ 1100 K
above the bottom of the potential curve, and the
vibrational separation between the ground and the first
excited vibrational level is $\sim$ 2175 K.
The rotational constant for the molecule is 
about 1.95 K.
We have verified that the
inclusion of the first rotational levels
($N$ = 0, 2, 4, 6) of $v$ =1 modifies our
results only within 0.5\%. 

Results are shown in Fig. 1 on a bilogarithmic scale.
The two compared models
provide nearly perfectly consistent results:
the curves are virtually indistinguishable, results of the rigid rotor
differing by at most 10 \% from the
complete calculation that allows the O$_2$
molecule to vibrate. We have seen how this small discrepancies can
be washed out by adjusting the short range potential
of the rigid rotor Hamiltonian. This artificial
"fine-tuning" is already a 
common practice in cold collision theory,
as it enables both an accurate fit to experimental
data, and predictive power \cite{vankokheiver02}.
Our results suggest that the main influence of vibration
might be absorbed into a similar fine-tuning,
at least until high-resolution data are available
that demand a more accurate model.

An exception occurs near a
resonance, where the extreme sensitivity of
phase shifts to details of the potential 
alters the lineshape slightly.
However, the overall agreement is quite 
good and demonstrates the adequacy of the rigid
rotor model in this case. The reason for this
is obvious: the attractive well depth of the He - O$_2$
interaction is only $\sim$ 40 K, so that the 
incident He atom does not have nearly enough energy
to excite vibration in the molecule, even as
a virtual excitation.
We expect this conclusion 
to hold generally in BGC, owing to the relatively
weak interaction of Helium with anything, and
also for O$_2$ - O$_2$ cold collisions 
\cite{avdbohPRA01} since the intermolecular 
well depth is only $\sim$ 200 K. In this case in fact,
the atom-atom exchange (which could be affected by
vibration) is unlikely to take place, because
at low temperatures the two oxygen
molecules do not get close enough
in the collision process. 

For many systems this separation of the
energy scales may no longer be the case,
and vibrations may play a more important role.
To study this influence,
the He - O$_2$ interaction potential
($V'$ in eq. (\ref{totalH})) has been made
artificially deeper by multiplying it by an arbitrary 
factor $\lambda$, ranging from 1 to 100. 
For $\lambda$ = 100, the well
depth of the three-body potential is approximately
twice as large as the lowest vibrational 
excitation energy of O$_2$.
We note here that increasing $\lambda$ has 
two effects: first, it makes vibrational resonances
energetically possible, and, second, introduces many
more rotational resonances, since rotational energy
splitting is much smaller than the vibrational one.
This is made intuitively clear in Fig. 2, where
a set of adiabatic potential curves
for $\cal J$ = 1
are displayed for a large value of the factor $\lambda$.

We will refer in the following to very low collision
energy (1 $\mu$K) because this allows us to include
only total ${\cal J}$ = 1 in our calculations,
thus reducing computational effort
\cite{volbohPRA02}. Figure 3 plots, 
both for rigid rotor and vibrating diatom, the
elastic and total inelastic rate constants for the same
transitions considered in Fig. 1
as a function of $\lambda$.
As the potential is made deeper (larger 
$\lambda$), new He - O$_2$
bound states appear, which show up as resonance-like
features in the figure.
Two different $\lambda$ ranges are displayed,
corresponding to two different physical situations.
In one range ($\lambda$ = 23 - 25), vibrationally
excited molecular states are not energetically accessible,
whereas in the range 90 - 91
vibrational resonances are accessible.
In the first case, the resonance pattern appears just
a little shifted going from one model to the other,
while in the second case, apart from a bigger
shift, some different features are
present in the rate constant trends.

However, the magnitudes and overall
patterns of the rate constants are
comparable in the two models.
This suggests that, as for the original
He - O$_2$ problem ($\lambda$ = 1), in modeling cold collisions
the rigid rotor Hamiltonian can be adjusted 
(by varying $\lambda$ in this simple case)
to nearly reproduce the results of the full
vibrating case.
We have indeed seen this in
the energy dependence of the
rate constants
for several different 
$\lambda$ values.

In conclusion, we find that the rigid rotor model
is very accurate for cold collisions of He with
O$_2$. 
This has been proven for this system,
but can be extended as well to stronger
interaction potentials,
as long as the energetic gain in the
three body interaction does not exceed the vibrational
excitation energy. Even in this case, the discrepancies
can be handled by fine-tuning the interaction,
at least until high-resolution data become available.

We notice that this is not true for interactions like
A + A$_2$, where A is for example an alkali atom.
In such systems, moreover complicated by exchange
effects between identical atoms, vibration can not be
neglected $a$ $priori$, not even for collision energies
tending to zero.

\acknowledgements
This work was supported by the National Science Foundation
and by NIST. A. V. acknowledges financial support from the
Universit\`a degli Studi di Perugia (Italy).

\begin{figure}
\caption{Elastic and inelastic rate constants relative to the entrance 
channel $| N\;J\;M_J \rangle$ = $| 0\;1\;1 \rangle$ in the
100$\mu$K - 10 K collision energy range. For each curve the final state
is indicated. Solid and dashed lines refer to the vibrating diatom
and rigid rotor models, respectively: on this scale, the two models 
can barely be distinguished.}
\end{figure}

\begin{figure}
\caption{Adiabatic curves for total angular momentum $\cal J$ = 1 
when the He - O$_2$ interaction potential
is boosted by a factor $\lambda$ = 90.5.
For this value of $\lambda$, the three-body well depth
is larger than the vibrational excitation, leading to the
possibility of new vibrational resonances. Not all the
rotational states included in the calculation are shown here
in order to preserve the clarity of the picture.}
\end{figure}

\begin{figure}
\caption{Elastic and inelastic rate constants 
for the entrance channel
$| 0\;1\;1 \rangle$ as a function of the scaling factor $\lambda$
for collision energy $E$ = 1 $\mu$K. Solid and dashed lines
refer to the vibrating diatom and rigid rotor models, respectively.}
\end{figure}


\begin{references}
\bibitem[*]{byline}  Email: bohn@murphy.colorado.edu

\bibitem{Fioretti} A. Fioretti {\it et al.}, Phys. Rev. Lett. {\bf 80},
4402 (1998). 

\bibitem{Knize} T. Takekoshi, B. M. Patterson, and R. J. Knize,
Phys. Rev. Lett. {\bf 81}, 5105 (1998); Phys. Rev. A {\bf 59}, R5 (1999).

\bibitem{Stwalley} A. N. Nikolov {\it et al.}, Phys. Rev. Lett.
{\bf 82}, 703 (1999).

\bibitem{Heinzen} R. Wynar {\it et al.}, Science {\bf 287}, 1016 (2000).

\bibitem{Paul} J. Weiner, V. S. Bagnato, S. Zilio, and P. S.
Julienne, Rev. Mod. Phys. {\bf 71}, 1 (1999).

\bibitem{Forrey} N. Balakrishnan, R. C. Forrey, and A. Dalgarno,
Phys. Rev. Lett. {\bf 80}, 3224 (1998); R. C. Forrey, V. Kharchenko,
N. Balakrishnan, and A. Dalgarno, Phys. Rev. A {\bf 59}, 2146 (1999);
R. C. Forrey, {\it et al.}, Phys. Rev. Lett. {\bf 82}, 2657 (1999).

\bibitem{baldalJPCA01} N. Balakrishnan and A. Dalgarno,
J. Phys. Chem. A {\bf 105}, 2348 (2001).

\bibitem{Doylemag} J. M. Doyle, B. Friedrich, J. Kim, and
D. Patterson, Phys. Rev. A {\bf 52}, 2515 (1995).

\bibitem{DoyleVO} J. D. Weinstein, {\it et al.}, J. Chem. Phys.
{\bf 109}, 2656 (1998).

\bibitem{DoyleCaH} J. D. Weinstein {\it et al.}, Nature {\bf 395},
148 (1998).

\bibitem{egoweipatfridoyPRA01} D. Egorov, J. D. Weinstein,
D. Patterson, B. Friedrich, and J. M. Doyle,
Phys. Rev. A {\bf 63}, 30501 (2001).

\bibitem{friJCSFT98} B. Friedrich {\it et al.}, J. Chem. Soc.,
Faraday Trans. {\bf 94}, 1783 (1998).

\bibitem{bohPRA00} J. L. Bohn, Phys. Rev. A. {\bf 62}, 32701 (2000).

\bibitem{avdbohPRA01} A. V. Avdeenkov and J. L. Bohn, Phys. Rev. A. 
{\bf 64}, 52703 (2001).

\bibitem{vanjonbetmeiPRA01} S. Y. T. van de Meerakker,
R. T. Jongma, H. L. Bethlem, and G. Meijer,
Phys. Rev. A. {\bf 64}, 41401 (2001).

\bibitem{volbohPRA02} A. Volpi and J. L. Bohn,
submitted to Phys. Rev. A, (2002).

\bibitem{grostrJCP00} G. C. Groenenboom and I. M. Struniewicz, J. Chem. Phys.
{\bf 113}, 9562 (2000).

\bibitem{Cyb} S. M. Cybulski {\it et al.}, J. Chem. Phys.
{\bf 104}, 7997 (1996).

\bibitem{Mizushima} M. Mizushima, {\it The Theory of Rotating
Diatomic Molecules}, (Wiley, New York, 1975), p. 170.

\bibitem{fremildeslurJCP70} R. S. Freund, T. A. Miller, 
D. De Santis, and A. Lurio,
J. Chem. Phys. {\bf 53}, 2290 (1970).

\bibitem{Cazzoli} G. Cazzoli and C. Degli Esposti, Chem. Phys. Lett.
{\bf 113}, 501 (1985).

\bibitem{Arthurs} A. M. Arthurs and A. Dalgarno, Proc. Roy. Soc. 
{\bf A256}, 540 (1960).

\bibitem{Child} M. S. Child, {\it Molecular Collision Theory}
(Mineola, Dover Publications, 1996), p. 100.

\bibitem{johJCP73} B. R. Johnson, J. Comp. Phys. {\bf 14},
445 (1973).

\bibitem{vankokheiver02} E. G. M. van Kempen, S. J. J. M. F
Kokkelmans, D. J. Heinzen, and B. J. Verhaar,
cond-mat/0110610 (2001).


\end{references}
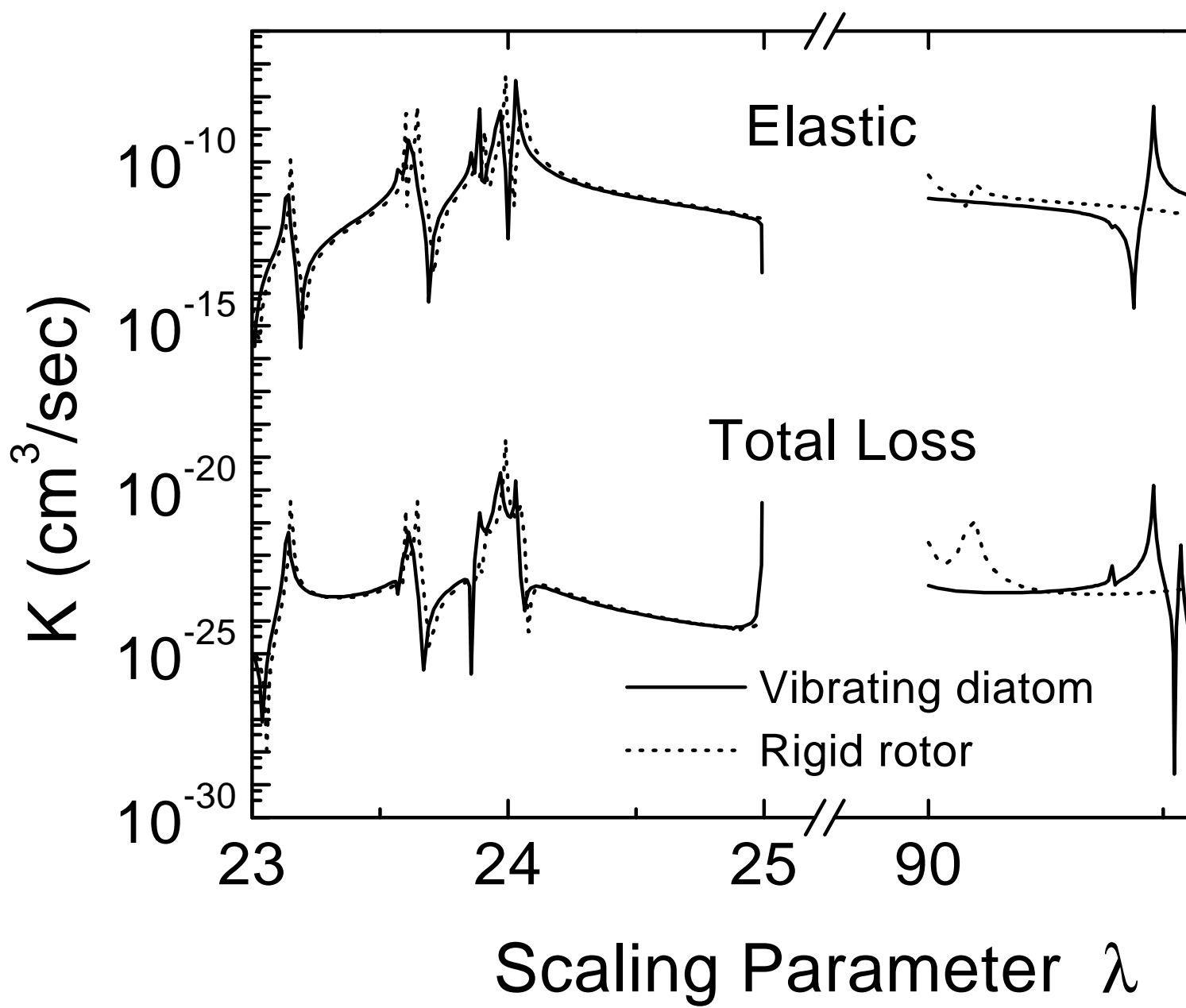
\end{document}